\documentclass[11pt]{revtex4}
\usepackage{graphicx}
\def\beq{\begin{equation}}
\def\eeq{\end{equation}}
\def\be{\begin{equation}}
\def\ee{\end{equation}}
\def\bea{\begin{eqnarray}}
\def\eea{\end{eqnarray}}
\def\nnb{\nonumber}

\newcommand{\gsim}{\lower.7ex\hbox{$\;\stackrel{\textstyle>}{\sim}\;$}}
\newcommand{\lsim}{\lower.7ex\hbox{$\;\stackrel{\textstyle<}{\sim}\;$}}

\begin{document}

 \title{ Radiation and energy release in a background field of axion-like dark matter}
 \author{ Wei Liao}
 \affiliation{
  Institute of Modern Physics, School of Science,
 East China University of Science and Technology,\\
 130 Meilong Road, Shanghai 200237, P.R. China
}


\begin{abstract}
 We find that a fuzzy dark matter background and the mG scale magnetic field in the galactic center can 
 give rise to a radiation with a very large energy release. The frequency of the radiation field is the same as 
 the frequency of the oscillating axion-like background field. We show that there is an energy transfer 
 between the fuzzy dark matter sector and the electromagnetic sector because of the presence of the 
 generated radiation field and the galactic magnetic field. The energy release rate of radiation is found to
 be very slow in comparison with the energy of fuzzy dark matter but could be significant comparing with the 
 energy of galactic magnetic field in the source region. Using this example, we show that the fuzzy dark matter 
 together with a large scale magnetic field is possible to give rise to fruitful physics.
\end{abstract}
 \maketitle

 \section{Introduction}\label{sec1}
An interesting hypothesis  of  dark matter(DM) in the universe is that the DM is composed of axion-like particles(ALPs).
For ALPs in different mass range, they may have very different behaviors in evolution of universe. For example, for a mass of ALP with $m_a \sim 10^{-5}$ eV,
ALPs can form boson stars ~\cite{Guth et al-2015}.  
For ultra-light ALP with $m_a \sim 10^{-23}-10^{-21}$ eV,  known as Fuzzy DM(FDM)~\cite{hbg-2000},
ALPs can have very large de Broglie wavelengths up to $1\sim 10$ pc and avoid typical problems associated
with Cold DM~\cite{hbg-2000, noCusp, Hui et al-2016}.
In this case, FDM is possible to form a diffuse DM background as a galactic halo~\cite{earlier, axionDM1}. 
In both of these cases, a very large number of light ALPs are concentrated in a volume of the scale of de Broglie wavelength.
Hence, physics of these ALPs can be described by a classical scalar field and quantum fluctuations around this classical field are small~\cite{Guth et al-2015}.

Because interaction of ALP with ordinary matter is very weak, detection of it in laboratory or observation of its cosmological or
astrophysical signature is very difficult~\cite{review}. Detecting the signatures of FDM is difficult in particular
because of its ultra-light mass scale. To date, there are very few ways known to us which can possibly constrain FDM~\cite{cosmo,KB}. 
In this Letter we discuss a possible physical prediction of FDM in the presence of a large scale magnetic field. 
Although our discussions are for FDM, the discussion can be easily extended
to ALP boson stars or condensates of other types of ALPs if there are magnetic fields with them.
 
 \vskip 0.5cm
\section{Electromagnetism in background of axion-like field} \label{sec2}
Considering coupling of ALP field $\phi({\vec x},t)$ with electromagnetic(EM) field, the Lagrangian can be written as
\bea
{\cal L}=-\frac{1}{4} F_{\mu \nu}F^{\mu \nu}+\frac{\phi}{2F} g_\phi {\widetilde F}_{\mu \nu} F^{\mu \nu}-A_\mu J^\mu \label{Lag}
\eea
where ${\widetilde F}_{\mu \nu}=\frac{1}{2}\varepsilon_{\mu \nu \rho \sigma} F^{\rho \sigma} $, 
$J^\mu$ being the 4-vector of an external current. 
$F$ in the denominator, known as the decay constant, is in the range $10^{16}\sim10^{18}$ GeV for FDM~\cite{Hui et al-2016}.
$g_\phi$ is a model-dependent parameter denoting the strength of the coupling of $\phi$ with
EM field. Its magnitude may vary from $10^{-3}$ to $10^{-2}$.
A modified set of equations of motion can be found using (\ref{Lag}). 
A $\phi F {\tilde F}$ term with a constant $\phi$ does not contribute to the equations of motion. 
However, in the presence of a background field $\phi({\vec x},t)$ which depends on space and time, 
the equation of motion with electric source becomes
\bea
\partial_\nu F^{\nu \mu} =J^\mu-g_\phi\frac{(\partial_\nu \phi)}{F}\varepsilon^{\mu \nu \rho \sigma} F_{\rho \sigma}. \label{Lag0}
\eea
Together with the equations with no magnetic source, we get a modified set of equations: 
\bea
&& {\vec \nabla}\cdot {\vec E}=J^0+g_\phi\frac{2}{F} ({\vec \nabla }\phi)\cdot {\vec B} \label{Eq1-1} \\
&& {\vec \nabla}\times {\vec B}-\frac{\partial {\vec E}}{\partial t}={\vec J}-g_\phi\frac{2}{F}[(\partial_t \phi) {\vec B}+({\vec \nabla}\phi)\times {\vec E}]
\label{Eq1-2} \\
&& {\vec \nabla}\cdot {\vec B}=0 \label{Eq1-3} \\
&&  {\vec \nabla}\times {\vec E}+\frac{\partial {\vec B}}{\partial t}=0. \label{Eq1-4}
\eea

For FDM in a galaxy, e.g. in Milky Way,  we can further simplify these equations.
Since the typical velocity in our galaxy is $\sim 10^{-3}$, FDM in the halo of a galaxy
can be considered non-relativistic and be described by a classical field as~\cite{KB}.
\bea
\phi({\vec x}, t)=A({\vec x}) \cos(m_a t+\beta({\vec x})) \label{PhiField}
\eea
The de Broglie wavelength of this non-relativistic FDM is found to be
\bea
\lambda_a=\frac{1}{m_a v_a}=63.6~\textrm{pc} ~\frac{10^{-22} ~\textrm{eV}}{m_a} ~\frac{10^{-3}}{v_a}. \label{wavelength}
\eea
This means that the spatial derivative of $\phi$ should be proportional to $1/\lambda_a$ and should be small
comparing with $\partial_t \phi$, i.e. $|{\vec v}|=|({\vec \nabla} \phi) / (\partial_t\phi)| \ll 1$.

So for FDM, we can neglect terms with ${\vec \nabla}\phi$ in Eqs. (\ref{Eq1-1}) and (\ref{Eq1-2}) as a first approximation,  
as long as we do not consider the case with strong electric field. We get
\bea
&& {\vec \nabla}\cdot {\vec E}=J^0 \label{Eq2-1} \\
&& {\vec \nabla}\times {\vec B}-\frac{\partial {\vec E}}{\partial t}={\vec J}-g_\phi\frac{2}{F}(\partial_t \phi) {\vec B}
\label{Eq2-2} \\
&& {\vec \nabla}\cdot {\vec B}=0 \label{Eq2-3} \\
&&  {\vec \nabla}\times {\vec E}+\frac{\partial {\vec B}}{\partial t}=0. \label{Eq2-4}
\eea
Note that in the approximation that  ${\vec \nabla}\phi$  is neglected, the right-handed side of (\ref{Eq2-2}) is indeed divergence free and
can be considered as an effective current of a neutral source with $J^0=0$:
${\vec J}_{eff}={\vec J}-g_\phi\frac{2}{F}(\partial_t \phi) {\vec B}$ and ${\vec \nabla}\cdot {\vec J}_{eff}=0$.

Neglecting the gravitational potential and the potential energy caused by axion interaction,
the energy-momentum tensor of FDM can be estimated using a free field Lagrangian of $\phi$~\cite{KB}.
One can find that the leading term in energy density is time independent
\bea
\rho_{DM}({\vec x})=\frac{1}{2}m^2_a A^2({\vec x}) \label{energydensity}
\eea
and the oscillating part in energy density is proportional to $({{\vec \nabla} \phi} )^2\propto  {\vec v }^{ 2} \rho_{DM}$ which can be
neglected.  The energy density should vary slowly within a de Broglie wavelength (\ref{wavelength})
and be taken as a constant within a distance smaller than the de Broglie wavelength $\lambda_a$.
In the following, we will use (\ref{energydensity}) and replace $m_a A({\vec x})$ with $\sqrt{2 \rho_{DM}({\vec x})}$.

Apparently, the modified set of equations (\ref{Eq1-1}$-$\ref{Eq1-4}) may have very rich physical consequences.
For example, one would expect that the cosmic or galactic ALP field may affect the evolution and development of
cosmic or galactic magnetic field.  In the present Letter we are not going to discuss this complicated problem.
We assume that galactic magnetic field has been generated and serves as an external B field,
and then discuss the possible radiation caused by the effective current coming from the galactic FDM field and
external galactic B field. 

\vskip 0.5cm
 \section{Energy release in oscillating FDM field and external B field}\label{sec3}
 Magnetic field in our galaxy, in particular in galactic center, has been measured by 
 astrophysical observations\cite{GalacticB, GalacticB1, Morris0,Morris}. 
 It was found that inside a ring-like Central Molecular Zone(CMZ) the magnetic
 field in the galactic center is mainly perpendicular to the galactic plane~\cite{Morris},
 in particular in regions close to non-thermal filaments(NTFs).
 The magnetic field  in the CMZ is mainly toroidal, that is oriented parallel to the galactic plane.
 The strength of magnetic field in the galactic center is quite uncertain. A global picture is that a pervasive
 magnetic field with a strength of mG exists in a central region with a radius no less than 150 pc.
 Challenges to this picture of pervasive magnetic field exists\cite{GalacticB, Morris}. 
 An alternative picture is that magnetic field in local regions of NTFs is of strength around mG and 
 it is of strength of tens $\mu$G in the diffuse inter-cloud region.
 In this Letter, we do not discuss the configuration and strength of the magnetic field in the galactic center,
 rather show that this large scale magnetic field together with the oscillating FDM field is very possible to give rise to
 radiation and large energy release. In particular, we focus on effects caused by poloidal magnetic field
 inside the CMZ.
 
We use simplified equations (\ref{Eq2-1}$-$\ref{Eq2-4}) to present our result. We denote the
external  B field as ${\vec  B}_{ex}$ and the electric and magnetic fields of radiation
as ${\vec E}_r$ and ${\vec B}_r$. We take the interaction of FDM with external B field as a perturbation,
and the radiation field also as a perturbation to the external magnetic field. Assuming the external B field satisfying a set of Maxwell
equations without interaction with FDM, we can get a set of equations for the radiation field to the first order as
\bea
&& {\vec \nabla}\cdot {\vec E}_r=0 \label{Eq3-1} \\
&& {\vec \nabla}\times {\vec B}_r-\frac{\partial {\vec E}_r}{\partial t}=-g_\phi \frac{2}{F}(\partial_t \phi) {\vec B}_{ex}
\label{Eq3-2} \\
&& {\vec \nabla}\cdot {\vec B}_r=0 \label{Eq3-3} \\
&&  {\vec \nabla}\times {\vec E}_r+\frac{\partial {\vec B}_r}{\partial t}=0. \label{Eq3-4}
\eea
 Writing ${\vec B}_r={\vec \nabla}\times {\vec A}_r$ and ${\vec E}_r=-\frac{\partial}{\partial t} {\vec A}_r$ ,
 we can get the propagation equation of ${\vec A}_r$ in radiation gauge which
 can be solved using a complex vector field ${\vec {\cal A} }_r$~\cite{Jackson}
 \bea
 &&{\vec A}_r({\vec x},t)=\textrm{Re}({\vec {\cal A}}_r), \label{solution1-1}\\
 && {\vec {\cal A}}_r({\vec x},t)=\frac{1}{4\pi} e^{-im_a t} \int d^3 y ~\frac{{\vec {\cal J}}({\vec y})}{|{\vec x}-{\vec y}|} 
 ~e^{ik |{\vec x}-{\vec y}|} \label{solution1-2}
 \eea
 where ${\vec {\cal J}}({\vec y})=g_\phi \frac{2i}{F}m_a A({\vec y}) {\vec B}_{ex}({\vec y}) e^{-i \beta({\vec y})}$.
 $k$ in (\ref{solution1-2}) is the wave number of radiation and we have $k=m_a$ in the present case with an oscillating FDM field.
 Introducing $ \lambda_k=2\pi/k$, we can find 
 \bea
 \lambda_k=0.4 ~\textrm{pc} ~\frac{10^{-22} ~\textrm{eV}}{m_a}. \label{lambda_k}
 \eea
$\lambda_k$ is much smaller than the spatial scale of the pervasive magnetic field in galactic center. 
 (\ref{solution1-1}) and (\ref{solution1-2}) suggest that there are periodic EM fields with a period
 \bea
 T=\frac{2\pi}{m_a}\approx 1.3 ~\textrm{year} ~\frac{10^{-22} ~\textrm{eV}}{m_a}\label{period}
 \eea
 in a fixed position in our galaxy.
 Effects of such a periodic EM field are very interesting subjects to study.

A real evaluation of  (\ref{solution1-2}) is very complicated and difficult, not only because we do not really know the detail of the magnetic
field in galactic center, but also because  $A({\vec y})$ and $\beta({\vec y})$ are uncertain.
Instead, we make a rough estimate of the radiation field and the radiation power.
In particular,  if there are special regions in galactic center in which the magnetic field is much stronger than surrounding regions,
e.g. in regions close to NTFs, 
we can evaluate the integration (\ref{solution1-2}) in these special regions  independently. So we can evaluate the contribution
of a particular source region to the radiation field. The total contribution of all these source regions can be obtained by summing their contributions.
This is the strategy of the rough estimate in the present Letter.
 
 To simplify calculation, we can evaluate radiation field at a distance far away from the source region
 and with $|{\vec x}| k \gg 1$, e.g. at the solar distance from the center of galaxy $R_\circ=8$ kpc. 
 For such a distance far away from a particular source region in the galactic center, we can find
\bea
&& {\vec B}_r= \textrm{Re}({\vec {\cal B}}_r ~e^{-im_a t+i kr}), ~{\vec E}_r= \textrm{Re}({\vec {\cal E}}_r ~e^{-im_a t+i kr}),\label{solution2-1} \\
&& {\vec {\cal B}}_r\approx-\frac{k}{2\pi}\frac{1}{r} g_\phi \frac{m_a A_S}{F} e^{-i\beta_S} \int_S d^3 y ~{\vec n}\times {\vec B}_{ex}({\vec y})
e^{-i k {\vec n}\cdot {\vec y}},\label{solution2-2} 
\eea 
where ${\vec n}={\vec x}/|{\vec x}|$ and $r=|{\vec x}|$. Electric field of radiation in the vacuum can be
estimated as ${\vec {\cal E}}_r=- {\vec n}\times {\vec {\cal B}}_r$.
$A_S$ and $\beta_S$ in (\ref{solution2-1}) and (\ref{solution2-2}) are the values of $A({\vec y}) $ and $\beta({\vec y})$ taken
at the source region. This replacement can be done 
because $e^{-i k{\vec n}\cdot {\vec y}}$  in (\ref{solution2-2}) is usually a function changing much faster than
$A({\vec y})$ and $\beta({\vec y})$. In our estimate, we restrict the integration to a source region with a scale no more than $\lambda_k$. 
 Writing $| \int_S d^3 y ~ {\vec  B}_{ex}({\vec y}) e^{-i k {\vec n}\cdot {\vec y}} | = | {\vec B}_{ex}| \Omega_S$ where
 $\Omega_S$ is the effective volume,  we can make an estimate of the order of magnitude.
 Suppressing angular dependence which does not affect the estimate of the order of magnitude, we find
\bea
|{\vec B}_r|\sim 6.7\times 10^{-9} ~\textrm{mG}~\bigg(\frac{10^{-22} 
~\textrm{eV}}{m_a} \bigg)^2 \frac{R_\circ}{r} \frac{g_\phi}{10^{-2}}\frac{10^{15} ~\textrm{GeV}}{F}
\bigg(\frac{\rho_{DM}(S)}{10^4 \rho_\odot}\bigg)^{1/2} \frac{|{\vec B}_{ex}|\Omega_S}{\textrm{mG}~\lambda_k^3}.
\label{radiationB}
\eea
where $\rho_\odot =0.3 $ GeV cm$^{-3}$ is the DM energy density at the position of solar system.
We have taken $10^{4} \rho_\odot$ as a reference energy density in calculation because the DM energy density can
increase to $10^3\sim 10^4$ of $\rho_\odot$  in the galactic center according to the popular
Navarro-Frenk-White(NFW) density profile of DM~\cite{NFW}.  For the estimate of $|E_r|$, the numerical factor
in (\ref{radiationB}) is replaced by $2.0 \times 10^{-7}$ V m$^{-1}$.

The time averaged radiation power  is estimated using (\ref{radiationB}) as
\bea
&&\frac{d P}{d\Omega}=\frac{1}{2} \textrm{Re}[r^2 {\vec n}\cdot({\vec E}_r\times {\vec B}_r)]  \nnb \\
&& \sim 1.0 \times 10^{39} ~\textrm{erg year$^{-1}$}
~\bigg(\frac{10^{-22} 
~\textrm{eV}}{m_a} \bigg)^4  \bigg( \frac{g_\phi}{10^{-2}}\bigg)^2 \bigg( \frac{10^{15} ~\textrm{GeV}}{F} \bigg)^2
\frac{\rho_{DM}(S)}{10^4 \rho_\odot} \bigg( \frac{|{\vec B}_{ex}|\Omega_S}{\textrm{mG}~\lambda_k^3} \bigg)^2.
\label{energyloss}
\eea
 One can compare (\ref{energyloss}) with $E_{DM}$ and $E_B$,
the typical total energy of DM and magnetic field in the source region of a volume $\lambda_k^3$. We can find
\bea
\frac{P}{E_{DM}}\sim 10^{-16} ~\textrm{year}^{-1}~\frac{10^{-22} ~\textrm{eV}}{m_a} 
\bigg( \frac{g_\phi}{10^{-2}}\bigg)^2 \bigg( \frac{10^{15} ~\textrm{GeV}}{F} \bigg)^2
\bigg(  \frac{|{\vec B}_{ex}|}{\textrm{mG}}\bigg)^2 \bigg( \frac{\Omega_S}{\lambda_k^3}\bigg)^2 \label{energyrelease1}
\eea
and 
\bea
\frac{P}{E_{B}}\sim 10^{-8} ~\textrm{year}^{-1}~\frac{10^{-22} ~\textrm{eV}}{m_a} 
\bigg( \frac{g_\phi}{10^{-2}}\bigg)^2 \bigg( \frac{10^{15} ~\textrm{GeV}}{F} \bigg)^2 \frac{\rho_{DM}(S)}{10^4 \rho_\odot}  \bigg(\frac{\Omega_S}{\lambda_k^3}\bigg)^2. \label{energyrelease2}
\eea
We see that the energy release rate is slow but may not be negligible.  
The energy release rate is really slow in comparison with the energy of DM in the source region.
However, the energy release rate
could still be significant comparing with the energy of the galactic magnetic field in the source region, in particular
if $g_\phi$ could be larger than $10^{-2}$.

To know in more detail about the energy
transfer in energy budget, we can use (\ref{Lag}) to get the equation of motion of $\phi$ in the presence of $\phi F {\tilde F}$ interaction
and find 
\bea
\partial_\mu T_\phi^{\mu \nu}=g_\phi \frac{1}{2F}(\partial^\nu \phi) {\widetilde F}^{\rho \sigma} F_{\rho \sigma}=
 -g_\phi \frac{2}{F}(\partial^\nu \phi) {\vec E}\cdot {\vec B}\label{energytransfer1}
\eea
where $T_\phi^{\mu \nu}=(\partial^\mu \phi) (\partial^\nu \phi)-\eta^{\mu \nu} \frac{1}{2}[(\partial \phi)^2 -m^2 \phi^2]$ 
is the energy-momentum tensor of $\phi$ with a free field Lagrangian.
Similarly, we can also use the modified Maxwell equation (\ref{Lag0}) to get the property of the energy-momentum
tensor for EM field.
Setting $J^\mu=0$  in  (\ref{Lag0}) and with a bit of algebra we find
\bea
\partial_\mu \Theta^{\mu \nu}_{em}= -g_\phi \frac{1}{2F}(\partial^\nu \phi) {\widetilde F}^{\rho \sigma} F_{\rho \sigma}=
 g_\phi \frac{2}{F}(\partial^\nu \phi) {\vec E}\cdot {\vec B},\label{energytransfer2}
\eea
where $\Theta_{em}^{\mu \nu}=-F^{\mu}_ {~~\lambda} F^{\nu \lambda} +\frac{1}{4}\eta^{\mu \nu} F^{\alpha \beta} F_{\alpha \beta}$
~\cite{Jackson}.  It is clear that there is an energy transfer between the FDM sector and EM sector.
The rate of energy transfer would be proportional to $(\partial_t \phi){\vec E}_r \cdot {\vec B}_{ex}$ in the present case.
One can show that there are an oscillating term with frequency $2 m_a$  and a constant term in $(\partial_t \phi) {\vec E}_r\cdot {\vec B}_{ex}$.
The constant term in $(\partial_t \phi){\vec E}_r\cdot {\vec B}_{ex}$ would give rise to a flow of energy
between the FDM sector and the EM sector. However, the direction of energy flow depends on
the sign of the imaginary part of $\int d^3x d^3y ~{\vec B}_{ex}({\vec x})\cdot {\vec B}_{ex}({\vec y}) e^{ik|{\vec x}-{\vec y}|}/|{\vec x}-{\vec y}|$
which depends on the detailed distribution of the magnetic field strength in the source region. If this double integration is positive in the source region,
energy is transferred from the FDM sector to the EM sector, and a rough estimate shows that the rate of energy injection to
EM sector would be of the same order of  magnitude of the energy loss rate of radiation shown in (\ref{energyrelease1}) or (\ref{energyrelease2}). 
Apparently, energy transfer between FDM sector and EM sector and the energy loss in radiation
should all be taken into account, e.g. in the development and evolution of galactic magnetic field.
A detailed study of this topic is beyond the scope of the present Letter.

As a comparison, we can also estimate the radiation caused by the Earth, the Sun and a magnetic neutron star.
For these stellar objects with scale much smaller than $\lambda_k$,  factor $e^{-ik {\vec n}\cdot k}$ in (\ref{solution2-2})
can be taken as one and we get $\int d^3 y ~{\vec B}_{ex}({\vec y})=\frac{2}{3}{\vec m}$ where
${\vec m}$ is the magnetic moment of the source object.  For magnetic moments of the Earth and the Sun
at order of $10^{22}$ A m$^2$ and $10^{29}$ A m$^2$, the radiation powers are 
$P \sim 10^{-14}$ erg year$^{-1}$ for the Earth and $P\sim 10^{-1}$ erg year$^{-1}$ for the Sun separately.
So they can be safely neglected.
For compact object like neutron star, the magnetic field
could be much more than $10^{10}$ G. However the magnetic moment of the neutron star is hard to be much larger
than that of a star like the Sun.  It is reasonable to expect  that the power of radiation caused by a magnetic neutron star
might be as large as that caused by the Sun but should not be much larger. 
We can see that the rate of energy release caused by these stellar objects are all very small.
On the other hand, the energy release caused by the large scale magnetic field in galactic center could be
huge.

We emphasize that results given in (\ref{energyloss}), (\ref{energyrelease1}) and (\ref{energyrelease2}) 
are for particular source regions in galactic center, not for the whole galactic center.
The total energy release given by the whole galactic center could be a sum of many such kind of sources
with strong magnetic field. This is in particular true for the possible picture that mG scale magnetic field comes only with NTFs.
So the rate of total energy release in the galactic center could be much larger than that given in (\ref{energyloss}).  
One should also notice that our estimate is very uncertain because our knowledge about the magnetic field in the galactic center is very uncertain.

\vskip  0.5cm
 \section{Conclusion}\label{sec4}
 In summary,  we have shown that the $\phi F {\tilde F}$ interaction term in an oscillating background field of axion-like DM and
 a background of magnetic field gives rise to an effective oscillating current in the Maxwell equation, 
 so that it can give rise to EM radiation and energy transfer between FDM and EM sectors.
 The frequency of the radiation field equals to the frequency of the oscillating background field of axion-like DM.
 In other words, the energy of the radiation photon equals to the energy of the non-relativistic axion-like DM, i.e.
 the mass of axion-like DM. In the case of FDM, the energy of the radiation photon is $10^{-23}\sim 10^{-21}$ eV
 and the period of the EM wave of radiation is $0.1\sim 10$ year.
 We have estimated the strength of radiation and find that the FDM background and the large scale magnetic field 
 in the galactic center can give rise to a radiation with a very large energy release. 
 We found that the energy release rate of radiation is very slow in comparison with the energy of FDM 
 but it could be significant comparing with the energy of galactic magnetic field in the source region.
 Needless to say, a very interesting question is how to directly detect this EM wave of a very long wavelength
 coming from the galactic center. Given the very weak strength of this radiation field at the solar distance from
 the galactic center, it should be a difficult and very challenging topic.

 We note that the radiation can be absorbed by plasma in galaxy, in particular by plasma in galactic center,
 so that it can possibly affect the physics of the development and evolution of plasma in galactic center.
 Detailed study of the energy release rate and the impact on galactic plasma can be done using models of galactic magnetic field 
 and models of plasma in galactic center. This detailed research is out of the scope of the present Letter.
 Other interesting physics include the energy transfer between FDM and EM sectors
in models of galactic magnetic field, the effect of oscillating FDM field in the
development and evolution of galactic, inter-galactic and cosmic magnetic fields, possible effects
caused by the effective charge in $({\vec \nabla}\phi)\cdot {\vec B}_{ex}$ term in (\ref{Eq1-1}),
effects of the predicted diffuse EM radiation field on the physics of CMB polarization and propagation of cosmic ray.
These topics are all of great interests for future study. In conclusion, we have pointed out
that the FDM background together with a large scale magnetic field can give rise to fruitful physics.

\begin{center} {\bf Acknowledgements} \end{center}
 This work is supported by National Science Foundation of
 China(NSFC), grant No. 11375065,  and Shanghai Key Laboratory
 of Particle Physics and Cosmology, grant No. 15DZ2272100. 
 I would like to thank ITP CAS Beijing for the hospitality during the course of this
 work.

\end{document}